\documentclass[a4paper, 11pt, fleqn]{article}
\usepackage{longtable}
\usepackage{graphicx}
\usepackage{amssymb,amsmath,epsfig,array}
\usepackage{multirow}

\begin{document}

\title{\bf{Spectroscopic Variability of the Compact Planetary Nebula Hb 12}}

\author{N.P. Ikonnikova$^1$\footnote{ikonnikova@gmail.com}, I.A. Shaposhnikov$^{1, 2}$, V.F. Esipov$^{\dag}$, M.A.
Burlak$^1$,\\
V.P. Arkhipova$^1$, A.V. Dodin$^1$, S.A. Potanin$^{1, 2}$, and
N.I. Shatsky$^1$}

\date{\it{$^{1}$Sternberg Astronomical Institute,
Moscow State University (SAI MSU),Moscow, 119234 Russia\\
$^{2}$Faculty of Physics,Moscow State University, Moscow, 119991
Russia }}

\renewcommand{\abstractname}{ }

\maketitle

\begin{abstract}

We present the results of our new low-resolution spectroscopic
observations of the young compact planetary nebula Hb 12 performed
in 2011--2020 with SAI MSU telescopes. We have measured the
intensities of more than 50 nebular emission lines in the spectral
range $\lambda$3687-9532, detected interstellar absorption
features, and conducted a search for absorptions belonging to the
possible secondary component of the central star. The extinction
coefficient has been estimated from the Balmer decrement to be
$c$(H$\beta$)=1.15$\pm$0.07. The distance has been found by
analyzing the interstellar extinction maps to be $D\approx2400$
pc. We have traced the history of the spectroscopic observations
of Hb 12, beginning with the first spectra taken by Aller (1951)
in 1945. We have detected a systematic increase in the relative
intensities of the nebular [O III] $\lambda$4959 and $\lambda$5007
lines and a decrease in the relative intensity of the auroral [O
III] $\lambda$4363 line, which has led to an increase in the
observed flux ratio $F(\lambda 4959+\lambda 5007)/F(\lambda 4363)$
by a factor of $\sim$4 from 1945 to the present time. The [O
III]/[O II] line ratio $F(\lambda 4363)/F(\lambda 3727+
\lambda3729)$ remains constant, suggesting that the degree of
ionization, on average, for the nebula is invariable. The
temperature of the exciting star has been estimated to be
$T\approx41~000$ K. We conclude that a decrease in the electron
temperature and, possibly, electron density in the [O III] line
formation region is mainly responsible for the spectroscopic
variability.

\end{abstract}

{\it {Keywords}}: planetary nebulae, spectroscopic variability, Hb
12, gaseous-envelope parameters, evolution.

\newpage

\section*{INTRODUCTION}

Planetary nebulae (PNe) are a product of the late evolution of
low- and moderate-mass ($M\sim0.8-8.0M_{\odot}$) stars. The
lifetime of such stars in the post-asymptotic giant branch
(post-AGB) stage of evolution depends on the stellar mass and the
mass loss rate on the asymptotic giant branch (AGB) and can range
from 100 to several thousand years. The interest in studying the
observational manifestations of the evolution of post-AGB stars
and young planetary nebulae has increased significantly in recent
years owing to the construction of new evolutionary models by
Miller Bertolami (2016), whose time scales are several-fold
shorter than those in previous models (Vassiliadis and Wood 1994;
Bl\"{o}cker 1995).

Several objects the change of whose emission-line spectra can be
caused by a rise in the temperature of the exciting star,
consistent with the idea of rapid evolution in the post-AGB stage,
have already been discovered. For example, the compact
low-excitation PN Hen 2-260 has shown a change of its spectrum in
the last 30 years: in the mid-1980s no nebular emission lines were
detected in its spectrum (Acker et al. 1991), in 2001 the [O III]
$\lambda$5007 line flux was 5 \% of the H$\beta$ flux (Escudero et
al. 2004), while in 2012 it was already about 7 \% (Hajduk et al.
2014). The authors of the latter paper believe that the
strengthening of nebular lines is related to the increase in the
degree of ionization in the nebula because of a rise in the
temperature of the exciting star.

There exist PNe whose spectroscopic variability is associated not
with the evolution of the central star, but with the change in
gaseous-envelope parameters due to a separate episode of enhanced
mass loss by the nonstationary PN nucleus. IC 4997, for which the
spectroscopic observations have been carried out for more than
half a century (Kostyakova and Arkhipova 2009; Arkhipova et al.
2020), may serve as a striking example of such PNe. What provoked
the stellar-wind strengthening, whether this was a single event,
and whether it will be repeated again such form or another remains
an open question.

Kondratyeva (2005) detected significant changes in two PNe, M~1-6
and M~1-11, based on 30--35 years of observations. In the opinion
of this author, a significant strengthening of the [O III] and
He~I lines and a weakening of the [S II] line (in M~1-6) suggest a
change of the physical conditions in the nebulae. It is reasonable
to assume that all of the observed effects were caused by an
increase in $T_{\text{eff}}$, but the available estimates of this
stellar parameter do not confirm this hypothesis. For M~1-11
sudden increases in $N_e$(S II), $T_e$(O III), and $T_e$(N II)
were recorded in 1996. The author assumes that these changes are
associated with some dynamic events in the nebula.

The goal of this paper was to check the possible spectroscopic
variability of the young compact PN Hb 12.

The PN Hb 12 (also known as PNG 111.8-02.8, VV 286, the Matryoshka
nebula) was discovered by Edwin Hubble 100 years ago (Hubble
1921). This object is characterized by a complex bipolar structure
and compact sizes: the angular diameter of the optically brightest
part of the nebula is $2''$-$3''$, the region of weak H$\alpha$
emission extends approximately to 13$''$ (Miranda and Solf 1989).
Kwok and Hsia (2007) studied the nebular structure in the [N II]
line based on Hubble Space Telescope observations. Subsequently,
Vaytet et al. (2009) created a morpho-kinematical model of Hb 12
based on imaging and high-resolution long-slit spectroscopy. The
authors for the first time revealed the presence of end caps (or
knots) in a deep [N II] $\lambda$6584 image of Hb 12 and measured
their radial velocities, $\sim$120 km/s. Clark et al. (2014)
presented near-infrared (IR) spectroscopic observations of Hb 12
using the Near-infrared Integral Field Spectrograph (NIFS) on
Gemini-North. Combining NIFS with the adaptive optics system
Altair, the authors studied in detail the inner structure of the
nebula.

The PN Hb 12 stands out for its low metallicity (Fa\'{u}ndez-Abans
and Maciel 1986; Perinotto 1991; Hyung and Aller 1996; Kwitter et
al. 2003) and belongs to the Galactic thin disk. Given its
chemical composition, location in the Galaxy, and kinematical
characteristics, Quireza et al. (2007) classified Hb 12 as a type
IIa PN according to Peimbert (1978).

The central star may be a close binary system with an orbital
period of 3.4 h (Hsia et al. 2006); the effective temperature of
the PN nucleus is estimated to be 42 000 K (Preite-Martinez and
Pottasch 1983).

The PN Hb 12 has a significant far-IR excess and was identified
with the IR source IRAS 23239+5754. Zhang and Kwok (1991)
constructed the spectral energy distribution of the object in a
wide wavelength range, from 0.1 to 100 $\mu$m, and ascribed the IR
excess to the radiation of dust with $T_{\text{dust}}=203$ K.
Subsequently, Jiang et al. (2013) analyzed the spectrum of Hb 12
based on observations from the Spitzer satellite with the IRC
spectrograph and detected a set of silicate emission features with
a predominance of enstatite (MgSiO$_3$) in the spectral range
9.9-37.2 $\mu$m.

The basic known data on the object are given in
Table~~\ref{tabl1}.

The history of the spectroscopic study of Hb 12 spans 75 years
since the time when L. Aller took the first spectrograms of the
nebula at the Lick Observatory on July 2, 1945. The measurements
of the relative intensities of 16 emission lines in the spectral
range from 3727 to 5007 \AA\ are presented in Aller (1951).

Hb 12, along with other PNs, fell into various spectrophotometric
surveys: the photoelectric photometry of 34 PNe (O'Dell 1963), the
photographic and photoelectric spectrophotometry of the blue part
of the optical spectral range for 21 nebulae (Kaler et al. 1976),
the photoelectric measurements of emission line intensities for 36
nebulae (Barker 1978), and the photoelectric spectrophotometry of
8 compact nebulae (Ahern 1978). Note also the paper of Kwitter et
al. (2003), where the spectra of 21 objects, including the nebula
Hb 12, were analyzed to study the relative abundances of various
elements in PNe.

The most comprehensive and detailed study of a high-resolution
spectrum for Hb 12 in the range 366-1005 nm was carried out by
Hyung and Aller (1996). In particular, these authors determined
the physical parameters and chemical composition of the nebula and
constructed its spatial model. Based on high-resolution spectra,
Miranda and Solf (1989) studied the kinematical and geometrical
structure of Hb 12.

Luhman and Rieke (1996) studied in detail the near-IR spectrum of
Hb 12. Subsequently, Hb 12 was among the 72 PNe for which
spectroscopic observations were performed in the wavelength range
2.5-5.0 $\mu$m with the Infrared Camera (IRC) onboard the AKARI
satellite (Ohsawa et al. 2016).

The important question about the possible PN nucleus binarity
remains open. The bipolar structure of the emitting PN envelope
implies the presence of a binary star system as a consequence of
the influence of its gravitational field on the gas expansion.
However, no other reliable confirmations of the existence of a
companion, except for the paper by Hsia et al. (2006), where the
possible binarity of the central star of Hb 12 was studied based
on photometric and spectroscopic observations, have been obtained
so far for Hb 12. The authors of the mentioned paper deem the
presence of a secondary component proven and even provide an
estimate of its mass and orbital parameters, but the results of
this study came under criticism in De Marco et al. (2008).

This paper is devoted to the analysis of long-term spectroscopic
observations of Hb 12 aiming to study the possible variability of
the object.


\begin{table}[h!]
    \centering
    \small
    \caption{Basic data on the PN Hb 12}
    \label{tabl1}
    \begin{tabular}{|c|c|c|}
    \hline Parameter & Value & Source \\
    \hline
    Equatorial coordinates (J2000) & $\alpha$ = $23^{\text{h}} 26^{\text{m}} 14.8^\text{{s}}$, & \multirow{2}{*}{SIMBAD} \\
    (J2000) & $\delta=+58^{\circ} 10' 54.5''$ & \\
     \hline
     Galactic coordinates (J2000) & $l=111.^{\circ}88$, & \multirow{2}{*}{SIMBAD} \\
     (J2000) & $b=-2.^{\circ}85$ & \\
     \hline  Radial velocity   & $\sim$ -5.0 km/s & SIMBAD \\
     \hline  Excitation class  & 4 & Hyung and Aller (1996) \\
     \hline  Characteristics   & $B=14.^{m}5$, $V=13.^{m}8$, $\log g=5.5$, & \multirow{2}{*}{Hyung and Aller (1996)} \\ of the central star  & $R=3.0 R_\odot$, $L=1200 L_\odot$ & \\
     \hline  Expansion velocity of the central region  &  $\sim$16 km/s & Miranda and Solf (1989)\\
     \hline  Kinematic age &  300 years & Miranda and Solf (1989)  \\
     \hline  Mass of the HII region  &  0.05 M$_\odot$ & Miranda and Solf (1989) \\
    \hline
    \end{tabular}
\end{table}

\section*{OBSERVATIONS AND DATA REDUCTION}

The spectroscopic observations of the PN Hb 12 in 2011--2019 were
carried out with the 1.25-m telescope at the Crimean Astronomical
Station (CAS) of the SAI MSU using a spectrograph with a 600
lines/mm diffraction grating and a long slit of width 4$''$ in the
range 4000--9000 \AA. The full recorded spectrum consists of
several overlapping ranges. The observations were carried out with
several exposure times. The minimum exposure time was chosen so
that the strongest lines (H$\alpha$ and [O III] $\lambda$5007)
were not overexposed. On the observing night we imaged a star, a
secondary spectrophotometric standard, at a close airmass together
with the object for the subsequent flux calibration of its
spectra. An ST-402 765$\times$510 pixel CCD (pixel size 9$\times$9
$\mu$m) was used as the detector. The spectral resolution (FWHM)
was 7.4 \AA.

In 2019 the instrumentation was upgraded, as a result of which a
Canon lens and a FLI PL-4022 2048$\times$2048 pixel CCD were
installed on the spectrograph instead of the Zeiss lens and the
ST-402 CCD. As a result, the vignetting virtually disappeared, the
simultaneously recorded spectral range expanded, the image quality
improved, and the signal-to-noise ratio increased, especially in
the blue part of the spectrum.

The spectra were reduced with the standard CCDOPS software and the
SPE software created at the Crimean Astrophysical Observatory
(Sergeev and Heisberger 1993). We performed absolute flux
calibration based on the spectra of a standard star from the
spectrophotometric catalogue by Glushneva et al. (1998) using data
from the atlas of standard stellar spectra by Pickles (1985).

In addition, in November 2019 and January 2020 we took spectra in
the range 3500--7500 \AA\ with the 2.5-m telescope at the Caucasus
Mountain Observatory (CMO) of the SAI MSU using a new
low-resolution double-beam spectrograph (TDS) equipped with volume
phase holographic gratings. Andor Newton 940P cameras with a E2V
CCD42-10 512$\times$2048 pixel arrays are used as the detectors. A
detailed description of the instrument is given in Potanin et al.
(2020). The observations were carried out with a long slit of
width 1.$''$0. The data reduction included the bias subtraction,
flat fielding, and dark current subtraction. Cosmic-ray hits were
removed from the two-dimensional spectrogram. To correct for the
detector spectral efficiency, atmospheric and optics transmission,
we observed the spectra of standards for which the absolute
spectral energy distribution was retrieved from the
library\footnote{\text{https://www.eso.org/sci/observing/tools/standards/spectra/stanlis.html}}.
The entire reduction was performed using self-developed Python
scripts.

A log of spectroscopic observations is presented in
Table~\ref{tabl2}.


\begin{table}
    \caption{Log of observations}

    \begin{tabular}{|c|c|c|c|c|c|}

     \hline
      Designation& Date &  JD & Number of frames & Exposure time, s & Standard \\
      \hline
      \multicolumn{6}{|c|}{CAS, Zeiss + ST-402} \\
      \hline
      11 & Aug. 26, 2011 & 2455800 & 21 & 60, 300, 900 & BS 8606 \\
      15 & Aug. 13, 2015 & 2457248 & 21 & 60, 600, 900, 1200 & 1 Cas \\
      16 & Oct. 5, 2016 & 2457667 & 19 & 30, 60, 600, 1200 & 12 Cas \\
      17 & Aug. 20, 2017 & 2457986 & 19 & 60, 300, 1200, 1800 & 12 Cas \\
      18 & Oct. 8, 2018 & 2458400 & 10 & 60, 300, 1800 & 12 Cas \\
      19 & July 26, 2019 & 2458690 & 24 & 30, 60, 120, 600, 1800 & BS 8606, HD 211073 \\
      \hline
      \multicolumn{6}{|c|}{CAS, Canon + FLI} \\
      \hline
      19a & Aug. 7, 2019 & 2450702 & 6 & 30, 300, 600, 1800 & BS 8606,\, $\tau$ Cas \\
      19b & Oct. 1, 2019 & 2458758 & 5 & 30, 60, 1800 & 4 Lac \\
      \hline
      \multicolumn{6}{|c|}{CMO} \\
      \hline
      19c & Nov. 7, 2019 & 2458795 & 26 & 5, 20, 60 & BS 8606 \\
      20 & Jan. 20, 2020 & 2458869 & 39 & 1, 3, 5, 30, 60, 300  & Hilt600  \\
      \hline
    \end{tabular}
    \label{tabl2}
\end{table}

The emission line fluxes were determined by two methods: spectral
profile integration and Gaussian fitting (this technique was
applied mainly to blends and strong lines). Initially, these
quantities were determined in absolute energy units; subsequently,
for convenience, they were converted to the relative scale of
$F$(H$\beta$)=100. If at one date of observations a specific line
was reliably measured simultaneously on several frames, then the
values obtained were averaged and the error was calculated as the
rms deviation from the mean in a given sample. The accuracy of
measuring the intensity of an individual line was about 10 \% for
the spectrograms taken in Crimea before the lens replacement,
about 7-8 \% after the lens replacement, and 5-6 \% for the CMO
spectra.

\section*{DESCRIPTION OF THE SPECTRUM}

\begin{figure}[h]
\centering
 \includegraphics[scale=1.05]{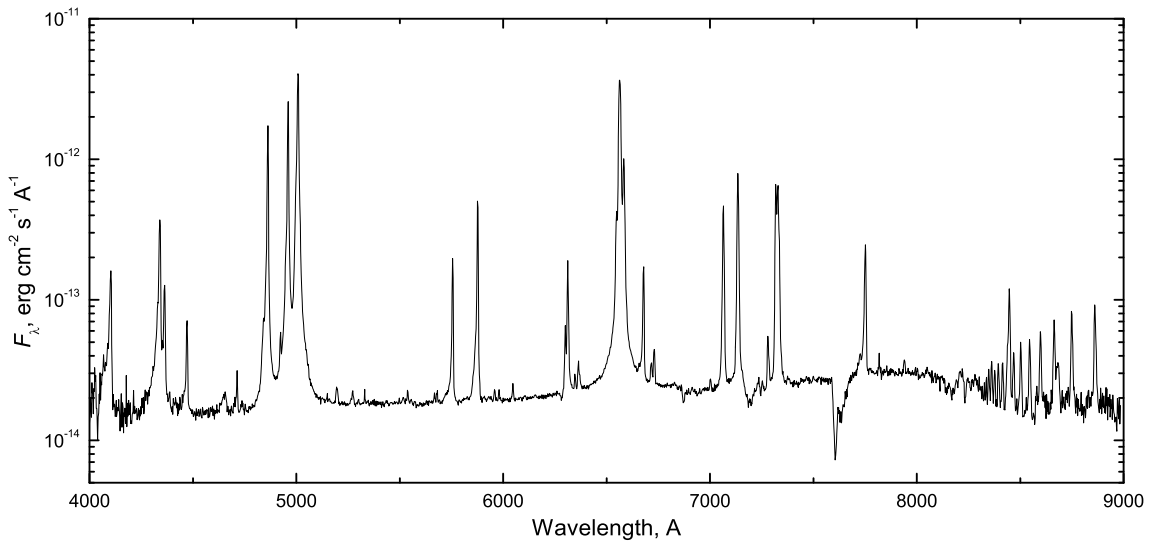}
  \caption{The CAS SAI spectrum of Hb 12 taken on August 13, 2015.}
 \label{fig1}
\end{figure}

For each set of CAS spectra we identified $\sim$50 emission lines
and measured their fluxes. The nebular [O III]  $\lambda$4959 and
$\lambda$5007, Balmer hydrogen lines, forbidden [O I], [O II], [O
III], [N II], [S II], [Ar III] lines, and He I recombination lines
are the most intense ones. In addition, there are permitted O I, O
II, N II, and N III lines in the spectrum. The Paschen and
forbidden [S III] lines (measured in the 2011, 2017, and 2018
spectra) are strong in the near infrared. In some spectra we
managed to identify and measure weak [Si II], [Cl III], [Fe III],
and some other ion lines, but the measurement errors of their
intensities are fairly large. An example of the spectrum taken in
Crimea is presented in Fig.~\ref{fig1}.


\begin{figure}[h!]
 \centering
 \includegraphics[scale=1.05]{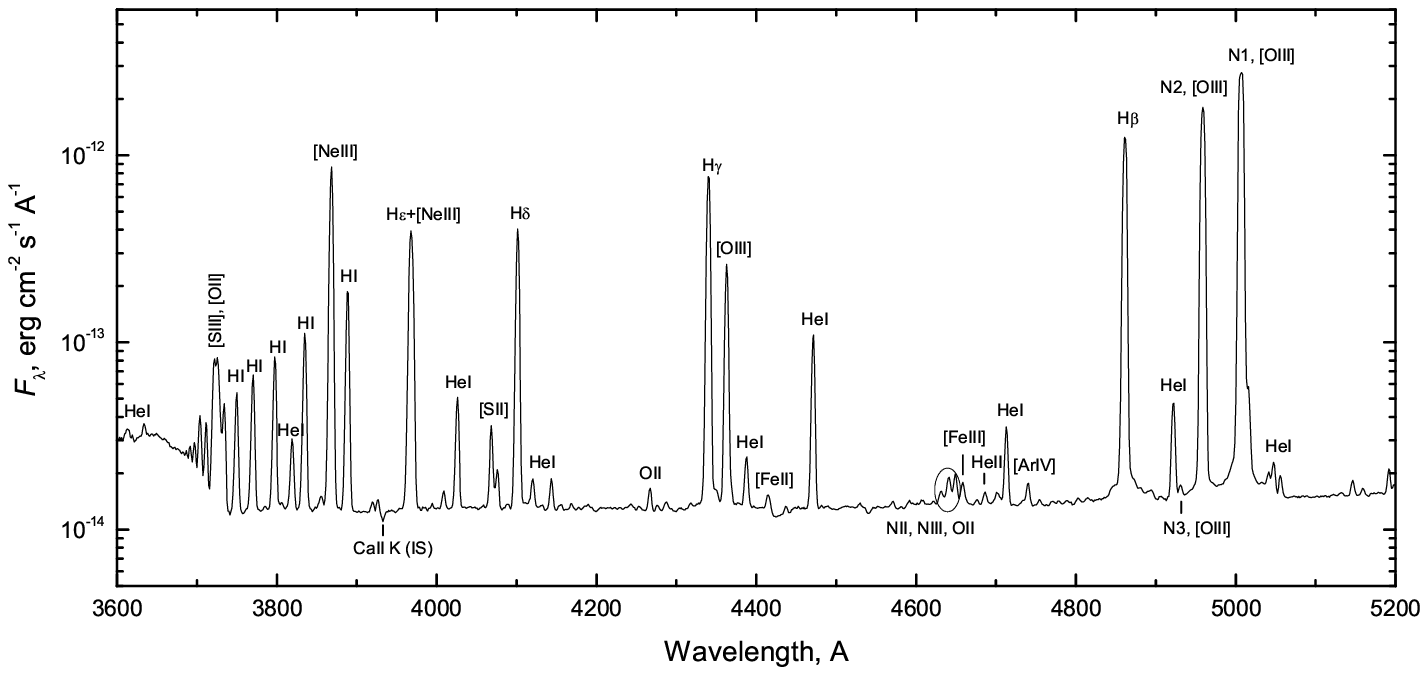}
  \caption{Fragment of the CMO spectrum for Hb 12 taken on January 20, 2020 (the wavelength range from 3600 to 5200 \AA).}
 \label{fig2}
\end{figure}

\begin{figure}[h!]
 \centering
 \includegraphics[scale=1.05]{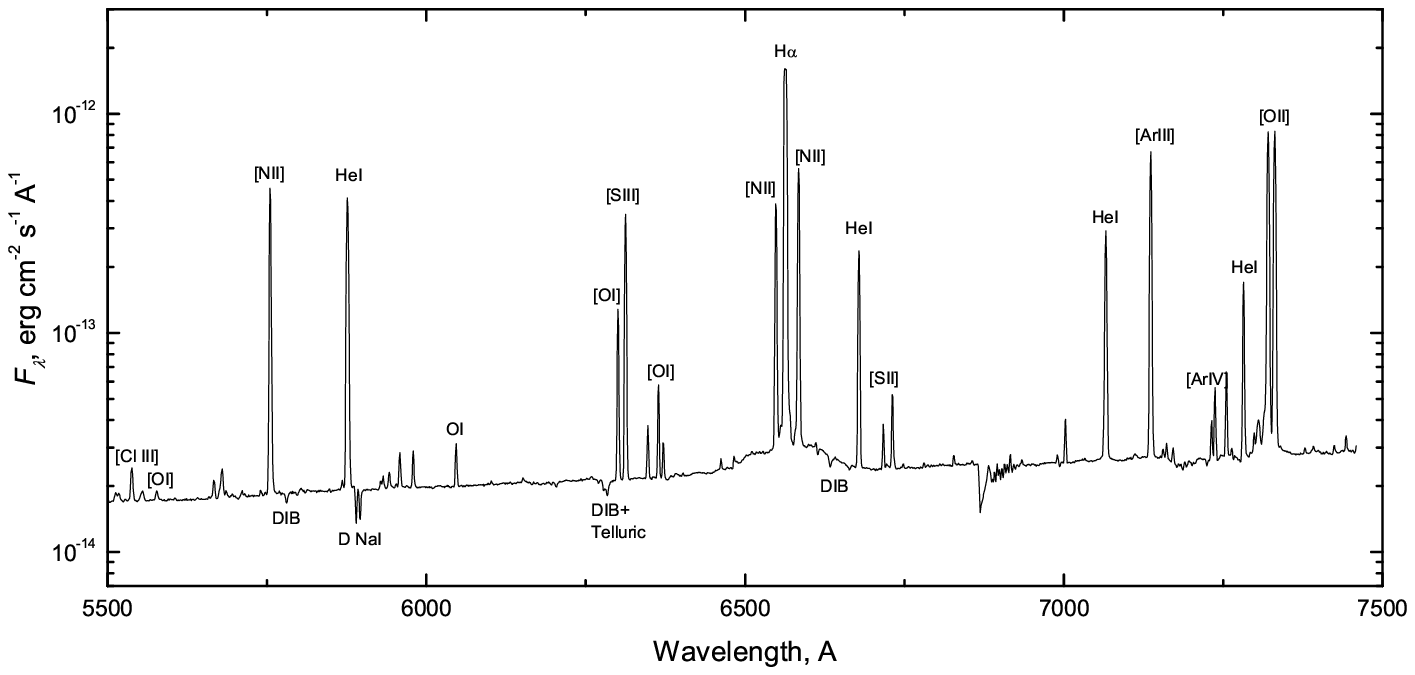}
  \caption{Fragment of the CMO spectrum for Hb 12 taken on January 20, 2020 (the wavelength range from 5500 to 7500 \AA).}
 \label{fig3}
\end{figure}

Compared to the technical capabilities of the CAS, the
instrumentation installed on the 2.5-m CMO telescope allows
spectra with a better resolution and a higher signal-to-noise
ratio to be obtained in a shorter exposure time. The blue and red
ranges are recorded simultaneously; the blue range was extended to
short wavelengths. Thus, new opportunities for a spectral analysis
open up. Figures~\ref{fig2} and ~\ref{fig3} present fragments of
the CMO spectrum in the wavelength ranges 3600--5200 \AA\ and
5500--7500~\AA, respectively.

In the CMO spectra in the blue part ($\lambda < 4000$ \AA) the
Balmer hydrogen lines and He I lines dominate and there are
permitted OI lines and forbidden [O II], [S III], and [Ne III]
lines. Based on these data, we also managed to measure the
intensity of the weak He II $\lambda$4686 line. At a wavelength
near $\lambda$4640 we see a group of N II, N III, and O II lines.
They probably belong to the central star and allow it to be
attributed to the PN cores with weak emissions in the spectrum
(the so-called \textit{wels}, weak emission-lines star) segregated
by Tylenda et al. (1993) into a separate subclass of central
stars.

The CMO spectrograms allowed us to measure the height of the
Balmer jump and to determine the electron temperature of the
nebula (see the corresponding section).

Apart from a rich line spectrum, we can see several absorption
features: the interstellar Ca II K (Fig.~\ref{fig2}), Na I D lines
and diffuse interstellar bands (DIBs) (Fig.~\ref{fig3}). In this
regard, the search for absorptions in the spectrum of Hb~12 that
may belong to the cool star, a probable companion of the central
one, is worth mentioning. Hsia et al. (2006) provided three
fragments of the spectrum for Hb~12 in which such absorptions were
identified. We made a comparison with our CMO data and found no
absorption features being discussed. In view of the better
resolution in the latter case, this suggests their absence
(except, possibly for the G band). Thus, the question of whether
the core of the nebula Hb~12 is binary remains open.

The spectroscopic data for Hb 12 obtained during the 2011--2020
observations are presented in the Appendix. The first two columns
give the wavelength (to within 1~\AA) and the corresponding
species. The F(year) columns give the observed (without any
correction for interstellar extinction) relative line intensities,
where by "year"\ we mean a separate data set (see the
"Designation"\ column in Table~\ref{tabl2}).

\section*{CHANGE OF THE SPECTRUM WITH TIME}

To keep track of how the spectrum of Hb 12 changed with time, we
used the data from Aller (1951), O'Dell (1963), Kaler et al.
(1976), Ahern (1978), Barker (1978), Hyung and Aller (1996),
Kwitter et al. (2003), and Hajduk et al. (2015) and compared them
with our observations performed at CAS in 2011-2019 and at CMO in
2019--2020.

Hb 12 is a compact PN: the size of its brightest central region
does not exceed 4$''$. The size of the slit or circular aperture
used in the observations is not always specified in the
literature. For those cases where this information is available,
obviously, the nebula entirely fell into the spectrograph slit.
This is also true for our observations performed in 2011--2019 at
CAS. In 2019--2020 the spectra were taken at CMO with a narrower
slit. However, our quasisimultaneous 2019 observations at CAS and
CMO showed that both the absolute H$\beta$ flux and the relative
emission line intensities are in good agreement, within the
measurement error limits. This gives us the right to include the
CMO data for 2019--2020 in our comparative analysis as well.

Table~\ref{tabl3} gives a compilation of the relative intensities
of some emission lines from Barker (1978), Hyung and Aller (1996),
and Kwitter et al. (2003), along with our new data. The CAS column
provides the averaged data obtained at the ZTE telescope in
2011--2019. The CMO column presents the relative line intensities
measured in the 2019--2020 spectra. Apart from these studies, we
took into account the data from all of the papers listed at the
beginning of the section when analyzing the changes in the
spectrum of Hb~12.


\begin{table}[h]
    \centering
    \small
    \caption{Observed relative intensities of emission lines in the spectrum of Hb 12 on the scale of
     $F(\textrm{H}\beta)=100$ and
     logarithm of the observed $\textrm{H}\beta$ flux in erg s$^{-1}$
     cm$^{-2}$}

    \begin{tabular}{|c|c|c|c|c|c|c|}
    \hline
   \multicolumn{2}{|c|}{\multirow{2}{*}{Source}}& \multirow{2}{*}{Barker 1978} & Hyung and & \multirow{2}{*}{Kwitter et al. 2003} & \multirow{2}{*}{CAS} & \multirow{2}{*}{CMO} \\
    \multicolumn{2}{|c|}{}&&Aller 1996&&&\\
    \hline
    \multicolumn{2}{|c|}{year}&1972&1990&1997&2011-2020&2019-2020\\
    \hline
    $\lambda$, \AA &Species&\multicolumn{5}{c|}{}\\
    \hline
   3727+29 & [O II] & 10.7 & 4.75 & 7.2 & - & 5.93 \\
   3869 & [Ne III]  & 38 & 38.28 & 34 & - & 33.9 \\
   4101 & H I   & 18.5 & 18.79 & 18 & 14.4& 15.4 \\
   4340& H I   & 33.9 & 35 & 33.4  & 34.3& 35.4 \\
   4363 &[O III]& 15.7 &  15 & 11.5 & 9.8 & 10.1\\
   4471 & He I  & 4.58 & 4.79 & 4.2 & 3.6& 4.5\\
   4686 &He II&0.3&0.07&--&--&0.14\\
   4861&H I&100&100&100&100&100\\
   4959 &[O III]& 144 & 168.27 & 160 & 185& 188 \\
   5007 &[O III]& 449 & 566.24 & 515 & 577& 601\\
   5192 &[Ar III]&--& 0.31 & --   & 0.41& 0.23\\
   5517 &[Cl III]&--& 0.08 & 0.1: &--& 0.08 \\
   5537) &[Cl III]&--& 0.21 & 0.2: & 0.31& 0.30 \\
   5755 &[N II]& 12.2 & 13.12 & 12.3 & 10.4& 12.0\\
   5876 &He I&29.6&  34.28&  30.7& 28.9& 31.6\\
   6300 & [O I]& \multirow{2}{*}{15} & 2.1  & 2.8 & 2.9& 2.7\\
   6312 & [S III] &  & 13.4 & 11  & 8.6& 9.0\\
   6363 & [O I]& 2.41 & 0.9  & 1.2 & 0.98& 0.88 \\
   6548 &[N II]&--   & 17.22 & 23.7 & -- & 20.1\\
   6563 & H I  & 752 & 1013.9 & 684 & 663& 861 \\
   6583 &[N II]& 68.1 & 60.81 & 63.9 & 74& 61\\
   6678& He I & 9.59 & 12.11 & 9.3  & 9.9& 11 \\
   6717 &[S II]& 0.96 & 0.3:  & 0.9  & 0.62& 0.44\\
   6731 & [S II]&1.57 & 1.0:  & 1.6  & 1.28& 0.88\\
   7065 & He I  & --  & 42.85   & 36.4 & 34& 38.7 \\
   7135 & [Ar III]&-- & 72.11   & 62.8 & 61& 62.6 \\
   7320+30& [O II]& 124 & 135.25 & 118 & 103& 113.7 \\
   7751 & [Ar III]&-- & 18.88  & 18 & 17.4& -- \\
   9069  & [S III]&-- & 124.17 & 117 & 105& -- \\
    \hline
    lg(F(H$\beta$)) & H I & -12.00$\pm$0.01 & -- & -10.96$\pm$0.04 & -10.95$\pm$0.02 & -10.91$\pm$0.03 \\
    \hline
    \end{tabular}
     \label{tabl3}
\end{table}

\begin{figure}[h!]
    \centering
    \includegraphics[scale=1.5]{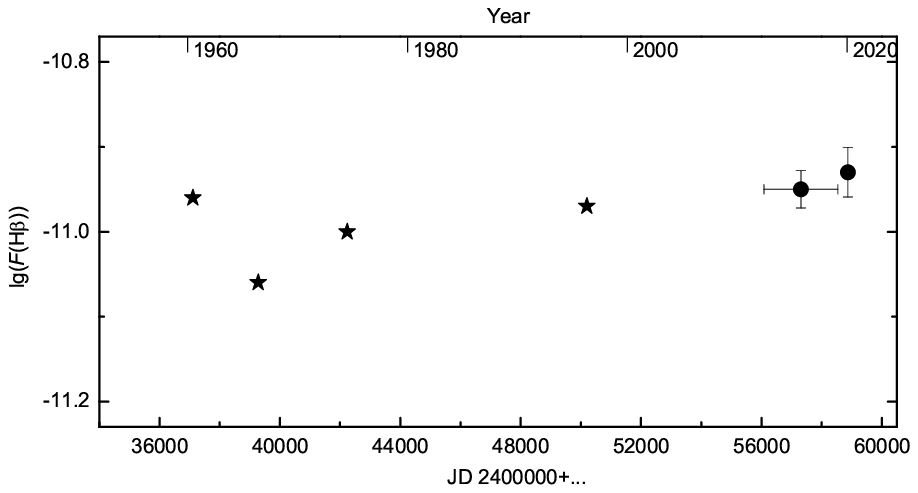}
    \caption{Time dependence of the logarithm of the H$\beta$ flux. The measurements from Barker (1978)
    (lg(F(H$\beta$))=--12) are not shown. The stars and the circles indicate the data from the
     literature and our measurements, respectively.}
    \label{fig4}
\end{figure}

Let us consider how the absolute H$\beta$ flux behaves with time
(Fig.~\ref{fig4}). On the whole, it can be said that the H$\beta$
flux did not change in the time of observations. However, some dip
that, if the measurement of Barker (1978) is taken into account,
could reach an order of magnitude is noticeable on the graph. It
exhibits no apparent correlation with the changes in the relative
intensities of other emission lines and, therefore, its nature
remains unclear.


\begin{figure}[h!]
     \centering
     \includegraphics[scale=1.5]{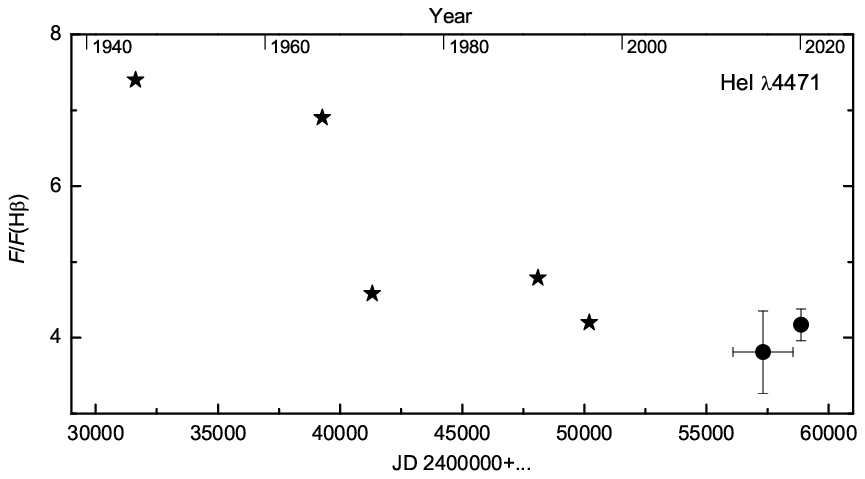}
     \caption{Change in the observed relative intensities of the He I $\lambda$4471 line with time.
     The stars and the circles indicate the data
     from the literature and our measurements, respectively.}
     \label{fig5}
\end{figure}

By analyzing the data presented in Table~\ref{tabl3}, we can
conclude that the relative intensity of the He I $\lambda$5876,
$\lambda$6678, and $\lambda$7065 lines has not changed
significantly over the period under consideration (from the first
half of the 1970s to the present time). The data for He I
$\lambda$4471 are available for a longer time interval, beginning
from 1945 (Aller 1951), and they suggest a possible decrease in
the relative intensity of this line (Fig.~\ref{fig5}). The He II
$\lambda$4686 emission line in the spectrum of Hb 12 is weak, and
it was measured only by Barker (1978) and Hyung and Aller (1996).
In the CMO spectra we managed to measure this line quite reliably.
As regards the behavior of He II $\lambda$4686, no changes outside
the error limits have been revealed so far in a short time
interval. It is expected that the He II line will strengthen as
the central star evolves.

\begin{figure}[h!]
     \centering
     \includegraphics[scale=1.5]{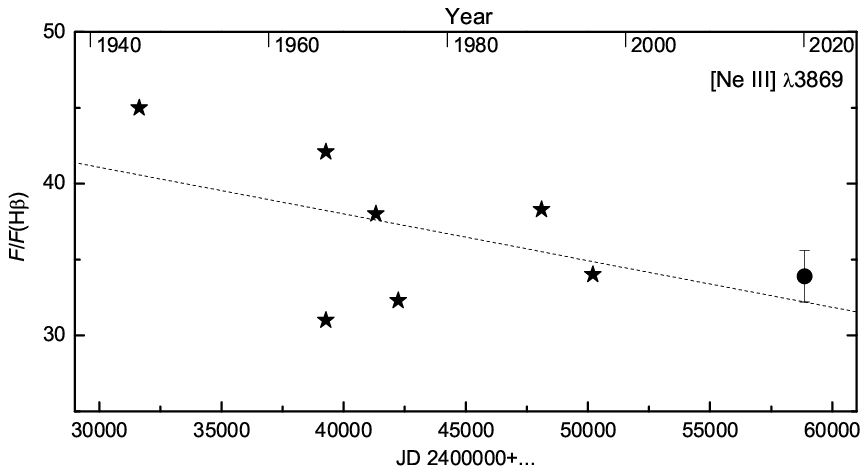}
     \caption{Change in the observed relative intensities of the [Ne III] $\lambda$3869 line with time. The stars and the circle indicate the
data from the literature and our measurement, respectively. The
dashed line represents the linear interpolation of all data.}
     \label{fig6}
\end{figure}

Consider the behavior of forbidden lines. The relative intensities
of the [Ar III] $\lambda$7135 and $\lambda$7751 lines and the
forbidden S$^{+2}$ $\lambda$6312 line apparently remain constant
so far. It is worth treating the possible time dependence of the
intensity for the forbidden [O I] $\lambda$6300 and $\lambda$6363
lines with caution because of the strong scatter in the data
obtained by different authors.

In the time of spectroscopic observations the relative intensity
of the [Ne III] $\lambda$3869 line (Fig.~\ref{fig6}) slightly
decreased, but these estimates also have a significant scatter.

\begin{figure}[h!]
     \centering
     \includegraphics[scale=1.5]{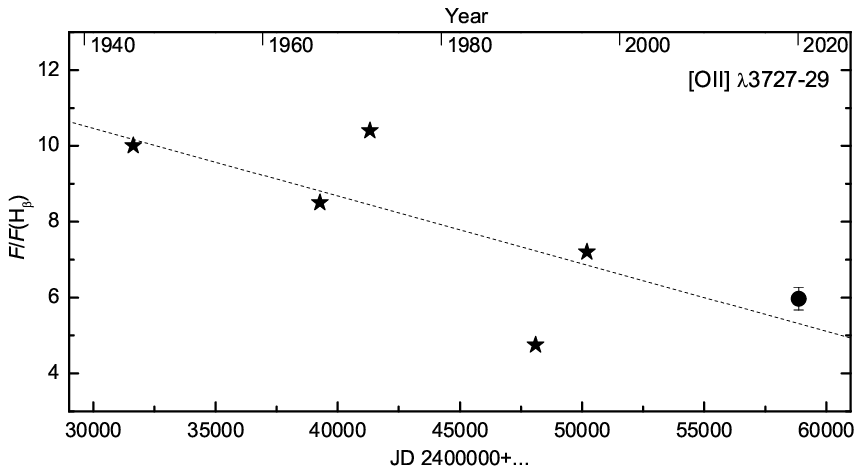}
     \caption{Change in the observed relative intensities of the [O II] doublet. The stars and the circle indicate the data from the
literature and our measurement, respectively. The dashed line
represents the linear interpolation of all data.}
     \label{fig7}
\end{figure}

The time dependence of the relative intensities of the [O II]
$\lambda$3727 and $\lambda$3729 doublet lines is plotted in
Fig.~\ref{fig7}. In view of the closeness of these lines in most
of the papers where the spectrum of Hb 12 was studied, including
our paper, the sum of these lines was measured. Despite the
significant scatter of data points, a falling trend can be
distinguished on the graph. The behavior of the IR [O~II]
$\lambda$7320 and $\lambda$7330 doublet also shows a similar
picture (see Table~\ref{tabl3}).


\begin{figure}[h!]
     \centering
     \includegraphics[scale=1.5]{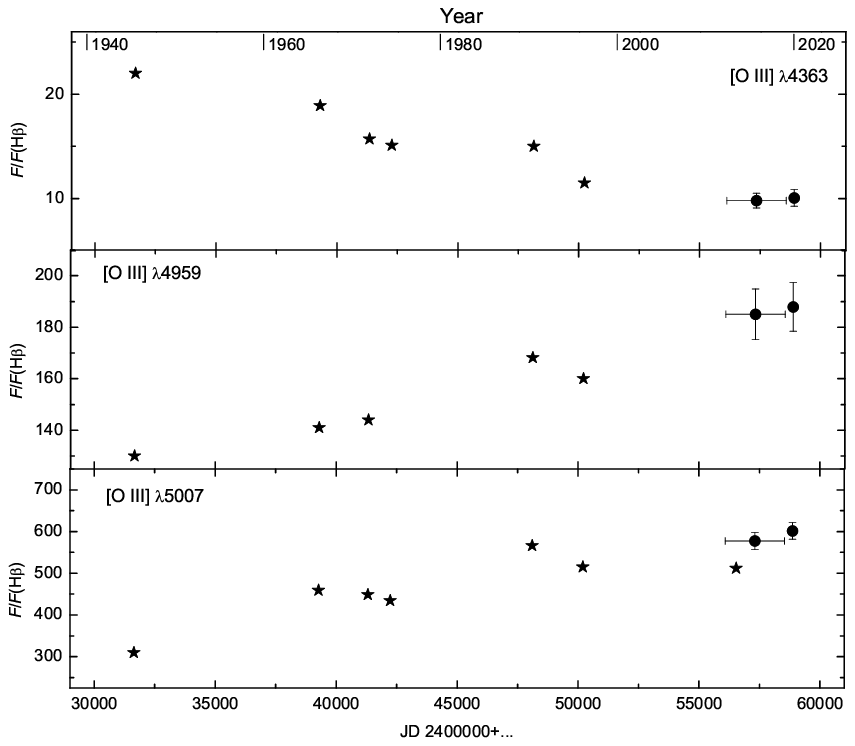}
     \caption{Change in the observed O$^{+2}$ line fluxes with time. The stars and the circles indicate the data from the literature and our
measurements, respectively.}
     \label{fig8}
\end{figure}

The behavior of forbidden [O III] lines, namely the nebular
$\lambda$5007 and $\lambda$4959 lines and the auroral
$\lambda$4363 line, deserves special attention. These lines are
among the strongest and reliably measured ones, and they are more
sensitive to a change in the parameters of the gaseous envelope
and the central star than others. Probably, there is, first, an
increase in the relative intensities of the $\lambda$4959 and
$\lambda$5007 lines and, second, a decrease in the ratio
$F(\lambda4363)/F(\mathrm{H}\beta)$ over the period under
consideration. Figure~\ref{fig8} plots the dependence of the
relative observed [O III] line fluxes on the epoch of
observations, which clearly illustrates this trend.

\section*{INTERSTELLAR EXTINCTION AND DISTANCE ESTIMATES}

In this paper the determination of the interstellar extinction was
based on a comparison of the theoretical relative intensities of
Balmer and Paschen hydrogen lines with the observed ones. The
extinction coefficient $c$(H$\beta)$ was determined using about
12--15 lines for each set of spectra or, more specifically, with
the application of optical Balmer lines for all spectra and
additionally ultraviolet lines of this series (for the CMO
spectra) or Paschen lines (for the CAS spectra). The theoretical
intensities of the Balmer and Paschen lines with respect to
H$\beta$ were taken from Hummer and Storey (1987) for the
following parameters: $T_e=10000$ K and $N_e=10000$ cm$^{-3}$.

The interstellar reddening curves $f(\lambda)$ were used in the
approximation from Cardelli et al. (1989), where $R$, the ratio of
total to selective extinction, was taken to be 3.2. The correction
was made using the well-known formula

\begin{center}
$\lg I(\lambda) - \lg F(\lambda) = c(H\beta)\cdot f(\lambda)$,
\end{center}

where $F(\lambda)$ and $I(\lambda)$ are the observed and corrected
line intensities, respectively, and $f(\lambda)$ denotes the
interstellar reddening law being used.

In other papers the interstellar extinction could be taken into
account by different methods: by directly determining the color
excess $E(B-V)$ by various techniques, by comparing the radio
continuum with the emission of hydrogen lines (Preite-Martinez and
Pottasch 1983), or using a larger or smaller number of Balmer and
Paschen lines. For example, Kwitter et al. (2003) made a
correction for reddening using only the ratio of H$\alpha$ to
H$\beta$. For comparison, here we provide the extinction
coefficients $c$(H$\beta$) for Hb 12 derived by other authors:
$c$(H$\beta$)=1.13 (Barker 1978), 1.25 (Rudy et al. 1993), 1.35
(Hyung and Aller 1996), and 1.05 (Kwitter et al. 2003).

Table~\ref{tabl4} gives the mean extinction coefficients
$c$(H$\beta$), along with the logarithm of the absolute H$\beta$
flux, for each date of observations. The mean $c$(H$\beta$) from
all our 2011-2020 data is $1.15\pm0.07$.


\begin{table}[h!]
    \centering
    \small
    \caption{The extinction coefficients and absolute fluxes in
the H$\beta$ line (in erg s$^{-1}$ cm$^{-2}$) calculated from the
results of each observation}
    \begin{tabular}{ccc}
      \hline
      JD &  c(H$\beta$)& $\lg$(F(H$\beta$)) \\
      \hline
       2455800  &   1.17& -10.90\\
       2457248  &   1.22& -10.92\\
       2457667  &   1.16& -10.94\\
       2457986  &   1.23& -10.95\\
       2458399  &   1.18& -10.92\\
       2458690  &   1.17& -10.97\\
       2450702  &   1.01& -10.98\\
       2458758  &   1.11& -10.95\\
       2458795  &   1.13& -10.91\\
       2458869  &   1.08& -10.94\\
      \hline
    \end{tabular}
    \label{tabl4}
\end{table}

\begin{figure}[h!]
  \centering
  \includegraphics[scale=1.1]{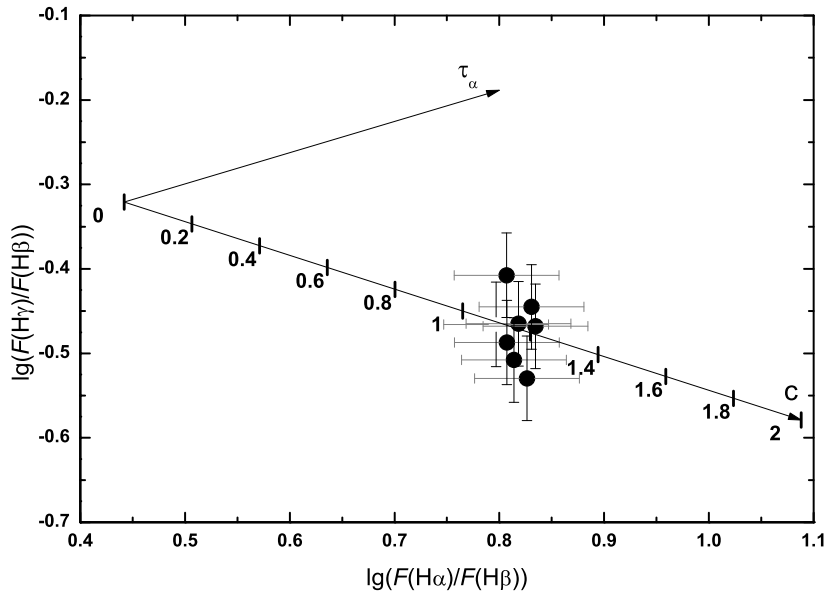}
  \caption{Logarithm of the ratio $F$(H$\gamma$)/$F$(H$\beta$) versus logarithm of the ratio
  $F$(H$\alpha$)/$F$(H$\beta$). The upper arrow specifies the
direction of displacement of the data points as the optical depth
in the H$\alpha$ line increases (Capriotti 1964); the lower arrow
corresponds to the interstellar reddening law from Seaton (1979).
The filled circles represent our results.}
 \label{fig9}
\end{figure}

Under conditions of a high gas density in the nebula (according to
the estimates by Hyung and Aller (1996), its value in the central
regions can reach $2\times10^6$~cm$^{-3}$), apart from the
calculation of the extinction coefficient proper, it is necessary
to study the question of possible self-absorption in hydrogen
lines, because this phenomenon can distort severely the result of
the correction for interstellar extinction. We followed the
procedure described in Burlak and Esipov (2009) as applied to the
PN IC~4997. In Fig.~\ref{fig9} the logarithm of the ratio
$F$(H$\gamma$)/$F$(H$\beta$) is plotted against the logarithm of
the ratio $F$(H$\alpha$)/$F$(H$\beta$). The values obtained for
each set of observations are plotted on the graph. It can be seen
that the data points are grouped near the line corresponding to
zero self-absorption, while their scatter does not exceed the
measurement error.


The distance estimation for the object is worth discussing
separately.

Determining the distances to PNe is known to be an open problem.
At present, no universal distance scale has been developed so far
for these objects. The various statistical and semiempirical
methods of determining the distances to PNe agree poorly between
themselves, and for some objects these estimates can differ by
several times.

Most of the Galactic PNe are extended objects, making it difficult
to measure the parallax and to calculate the distance by the
classical geometrical method. Nevertheless, PN researchers pin
great hopes on the Gaia mission (Brown et al. 2018). The first
results have already been obtained. For example,
Gonz\'{a}lez-Santamar\'\i a et al. (2019) derived reliable
distances for 211 objects from a total sample of 1571 PNe with
Gaia DR2 parallaxes, while Chornay and Walton (2021) presented an
updated catalogue of distances for PNe based on Gaia EDR3 data
(Brown et al. 2020). For the PN Hb 12 the parallax has not yet
been measured.

Yet another promising method of distance determination is based on
the knowledge of the interstellar extinction toward an object. The
availability of new photometric surveys, such as IPHAS (INT/WFC
photometric H$\alpha$ survey) (Drew et al. 2005), makes it
possible to use the so-called extinction method to determine the
distances to a large number of objects. Giammanco et al. (2011)
applied this method to a sample of 137 PNe. These authors
carefully studied the characteristics of the method and the main
sources of errors. The data available in the literature
supplemented with new observations allowed the distances for 70
PNe to be determined. For Hb 12 a fairly uncertain estimate of
$D<1000$ pc was obtained in this paper.

We decided to estimate the distance to Hb 12, knowing the
extinction coefficient for the nebula and using the interstellar
extinction maps constructed from the data of several extensive
surveys (Green et al. 2019). These maps allow the distance to the
object, depending on the color excess $E(g-r)$ in the photometric
SDSS system, to be determined. We converted the color excess
$E(g-r)$ to the total extinction in the $V$ band as
$A_V=(E(g-r)-0.03)/0.269$ (Green et al. 2019) and then to the
extinction coefficient using the formula
$c$(H$\beta$)=1.46$A_V/R$, where $R=3.2$.

\begin{figure}[h!]
  \centering
  \includegraphics[scale=1.5]{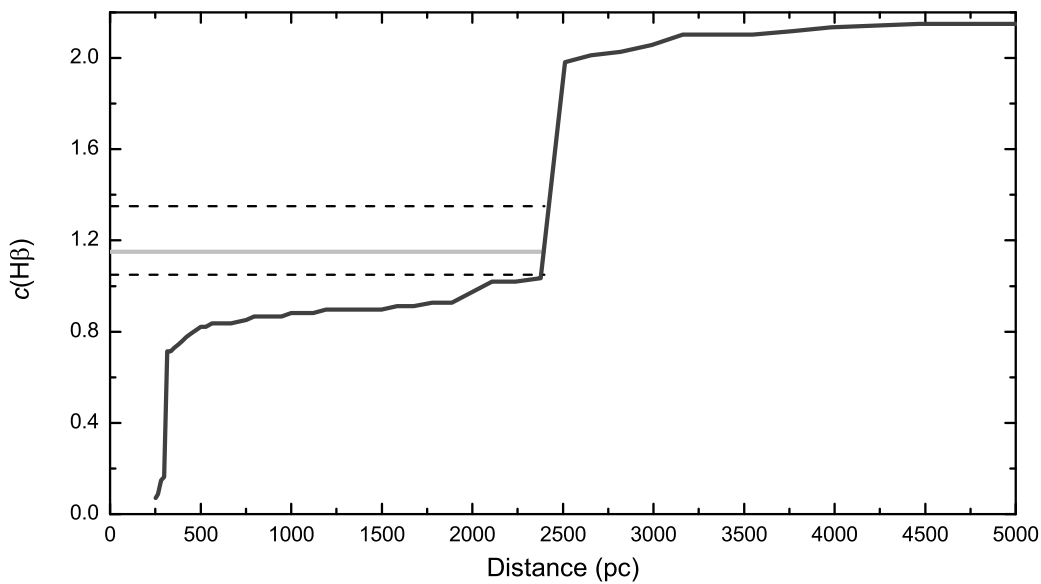}
  \caption{Extinction coefficient versus distance toward Hb 12 from the data
of Green et al. (2019) (black line). The gray line corresponds to
the mean extinction coefficient $c$(H$\beta$)=1.15 from our data.
The dashed lines bound the range of $c$(H$\beta$) from the
literature.}
 \label{fig10}
\end{figure}

In Fig.~\ref{fig10} $c$(H$\beta$) is plotted against the distance
toward Hb 12; the mean  $c$(H$\beta$)=1.15 from our data and the
$c$(H$\beta$) range from 1.05 (Kwitter et al. 2003) to 1.35 (Hyung
and Aller 1996) from the literature are shown. Let us compare the
distance $D\approx2400$ pc derived by us with the data of other
authors. The distance to Hb 12 has been determined repeatedly.
Table~\ref{tabl5} presents some estimates with an indication of
their source.


\begin{table}[h!]
    \centering
     \caption{Estimates of the distance to Hb 12 from the
literature}
    \begin{tabular}{c|c}
      \hline
      $D$, pc & Source\\
      \hline
      6700& Johnson et al. (1979)\\
      3030& Kingsburg and Barlow (1992)\\
      2236& Cahn et al. (1992)\\
      8110& Zhang (1993)\\
      2880& Phillips (2002)\\
      10460& Phillips (2004)\\
      2260$\pm$680& Frew et al. (2016)\\
      \hline
    \end{tabular}

    \label{tabl5}
\end{table}


Our $D$ estimate is closest to the estimates by Cahn et al. (1992)
and Frew et al. (2016). It can be seen from Fig. 10 that toward Hb
12 there is a sharp increase in extinction at a distance $>2500$
pc attributable to the presence of an interstellar cloud of gas
and dust. Therefore, the distance to the PN Hb 12 cannot exceed
2500 pc, and its values given in Kingsburg and Barlow (1992) and,
especially, Johnson et al. (1979), Zhang (1993), and Phillips
(2004) should be deemed overestimated.

\section*{DISCUSSION}

Thus,we have undoubted changes in the spectrum of Hb 12 over the
entire history of its spectroscopic observations. Let us now try
to interpret them based on the data obtained.

Initially, it was expected that the changes in the emission-line
spectrum of the nebula could be caused by the evolution of the
central star. In a number of papers the nebular O$^{+2}$ lines act
as the main indicator of a change in the central star temperature.
For example, Hajduk et al. (2015) studied the evolutionary changes
in the relative intensity of the [O III]  $\lambda$5007 line in
the spectra of 20 PNe. For some objects these authors detected an
increase in $F(\lambda5007)$/$F$(H$\beta$) with time and
associated it with a rise in the temperature of the ionizing
source.

For Hb 12 we detected an increase in the ratio
$F(\lambda5007)$/$F$(H$\beta$) by a factor of $\sim1.9$  in 75
years.

For optically thick low-excitation PNe Kaler (1978) proposed
empirical relations between the relative line intensities
$I(\lambda5007)$[O III]/$I$(H$\beta$) and $I(\lambda3869)$[Ne
III]/$I$(H$\beta$) and the temperature of the central star. We
compared the temperatures derived from the proposed formulas by
taking the data from this paper and the archival data from Aller
(1951) as a basis. The picture turns out to be contradictory: for
the [O III] line the temperature determined by this method shows a
rise in 75 years from $T_*\approx35~000$ K to $T_*\approx44~000$
K, while the values of $I(\lambda3869)$[Ne III]/$I$(H$\beta$) lead
to higher estimates of the temperature and correspond to its
decrease from $T_*\approx56~000$ K in 1945 to $T_*\approx49~500$ K
in 2020.

In view of this contradiction, we additionally determined the
temperature of the central star by the energy balance method
(Preite-Martinez and Pottasch 1983; Pottasch 1987). This method is
based on the idea of energy balance between the stellar radiation
and the surrounding gaseous nebula and requires calculating the
intensities of all the collisionally excited emission lines in the
nebular spectrum. Based on our data, we found the temperature to
be $\sim$41 000 K, consistent with the estimate of 42 000 K from
Preite-Martinez and Pottasch (1983) obtained almost 40 years ago.
Thus, the rise in the temperature of the central star in the time
of spectroscopic studies of Hb 12 is still open to question.

\begin{figure}[h!]
    \centering
    \includegraphics[scale=1.45]{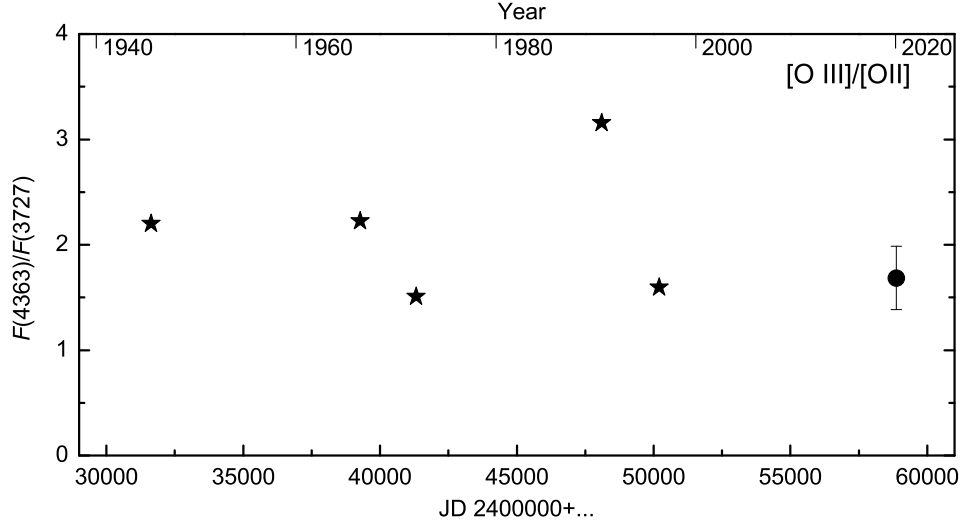}
    \caption{Intensity ratio of the [O III] $\lambda$4363 and [O II] ($\lambda$3727 + $\lambda$3729) lines. The stars and the circle
    indicate the data from the literature and our measurement, respectively.}
    \label{fig11}
\end{figure}

The [O III] $\lambda$4363/[O II] $\lambda$3727 intensity ratio,
which characterizes the degree of ionization averaged over the
nebula, lends support to the constancy of the temperature of the
ionizing source. For Hb 12 its value probably has remained without
any change over the last 75 years, while the scatter of data is
due to the observational errors and the inhomogeneity of the
observational data of different authors (Fig.~\ref{fig11}).


A change in gaseous-envelope parameters seems a more obvious cause
of the spectroscopic variability. Consider the most characteristic
line intensity ratios that reflect the physical conditions in the
nebula.

For the high-excitation zone of Hb 12 Hyung and Aller (1996) found
the parameters to be $N_e=5\times 10^5$ cm$^{-3}$ and $T_e=13~600$
K. As was shown by Ahern (1975), in the case of a high electron
density ($N_e>10^5$ cm$^{-3}$), the intensity ratio of the nebular
[Ne III] $\lambda$3869 and [O III] $\lambda$5007 lines depends
weakly on $T_e$ in the range 10 000-20 000 K and, therefore, can
serve as an $N_e$ indicator.

\begin{figure}[h!]
    \centering
    \includegraphics[scale=1.45]{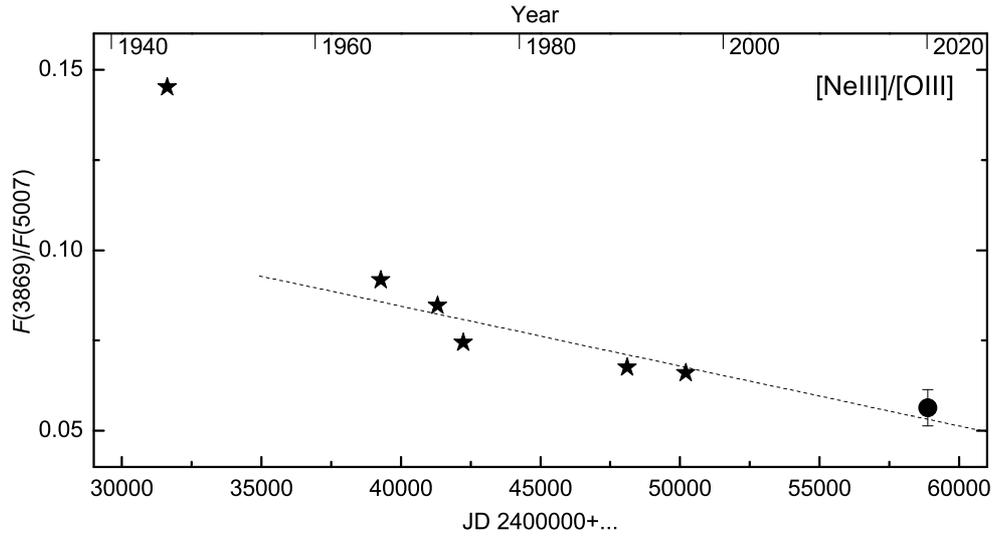}
    \caption{Intensity ratio of the [Ne III] $\lambda$3869 and [O III] $\lambda$5007 lines.
    The stars and the circle indicate the data from the literature
    and our measurement, respectively. The dashed line is the linear
    interpolation of the data without including the measurements
    of Aller (1951).}
    \label{fig12}
\end{figure}

Figure~\ref{fig12} shows the change in the intensity ratio of
these lines with time from the published data and our new
observations. From the first observations in 1945 to the mid-1960s
$F(\lambda 3869)/F(\lambda 5007)$ decreased by a factor of
$\sim$1.6; in succeeding years the drop in this ratio was not so
steep and the data are satisfactorily described by a linear
equation.


Figure~2 in Ahern (1975) presents the theoretical dependences of
$\lg(I(\lambda3869)/I(\lambda5007))$ on $\lg N_e$ for the mean ion
abundance $N(\textrm{NIII})/N(\textrm{OIII})$=0.22 for the PN.

\begin{figure}[h!]
    \centering
    \includegraphics[scale=1.2]{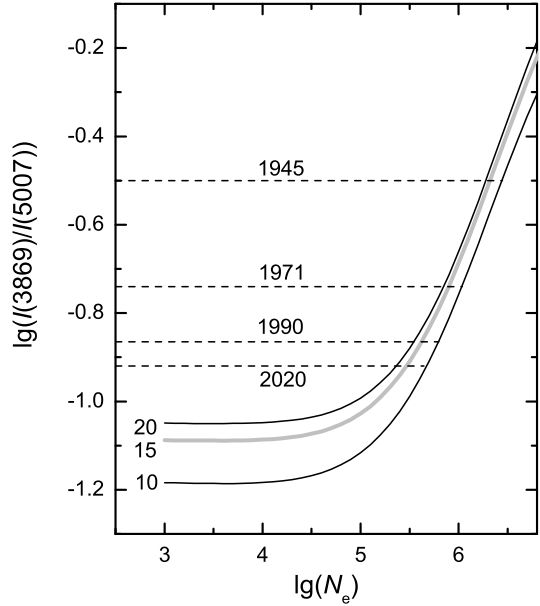}
    \caption{Logarithm of the ratio $I(\lambda3869)/I(\lambda5007)$ versus
    electron density. The 1945, 1971, 1990, and 2020 data are those
    from Aller (1951), Barker (1976), Hyung and Aller (1996), and this
    paper, respectively.}
    \label{fig13}
\end{figure}

We constructed such a diagram (Fig.~\ref{fig13}) at some fixed
temperatures (marked by the numbers in units of 1000 K) using the
emissivities from the Nebulio database mentioned in Giannini et
al. (2015). To construct the theoretical curves, we took the ion
abundance ratio $N(\textrm{NIII})/N(\textrm{OIII})$=0.184 based on
the data for Hb 12 from Hyung and Aller (1996).


In Fig.~\ref{fig13} we plotted the extinction-corrected 1945,
1971, 1990, and 2020 data referring to Hb 12 and estimated $N_e$
for different epochs. For example, if we take the range
$T_e=10~000-15~000$ K for Hb 12, then for the first epoch $\lg N_e
\simeq 6.3-6.4$, while for 2020 $\lg N_e \simeq 5.5-5.7$,
suggesting a drop in the electron density by a factor of $\sim$5
in 75 years. Such a significant change in $N_e$ is difficult to
explain by the gaseous-envelope expansion alone. At an expansion
velocity of the nebula $V_{\textrm{exp}}\sim16$ km/s (Miranda and
Solf 1989) its size increased by 25 \% in 75 years, which will
lead to a decrease in the electron density by no more than 60 \%.
However, it should be noted that the central part of the nebula
has a complex structure and also includes high-velocity bipolar
outflows (Clark et al. 2014).

Consider the diagnostic ratio of the nebular and auroral O$^{+2}$
lines.

\begin{figure}[h!]
    \centering
    \includegraphics[scale=1.45]{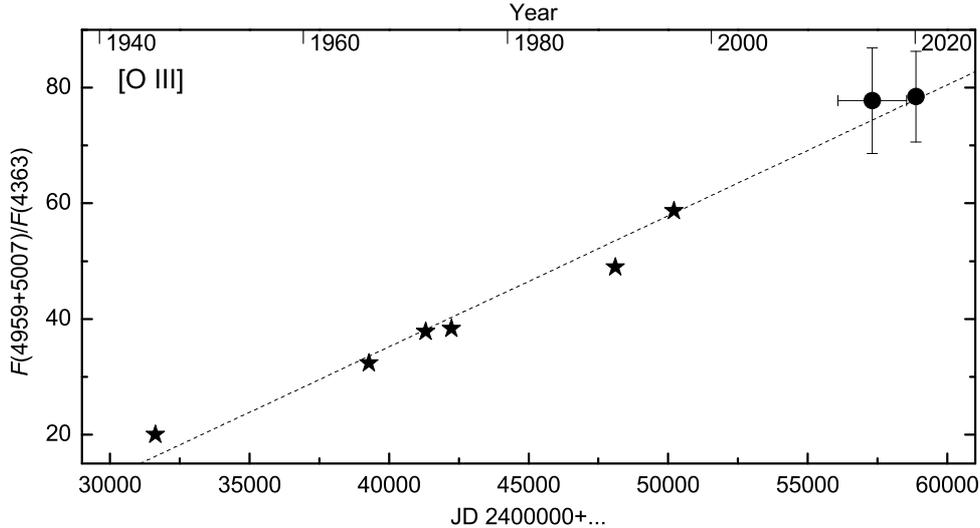}
    \caption{Time dependence of the ratio of the observed [O III] line fluxes: the stars and the circles indicate the data from the
literature and our measurements, respectively. The dashed line
represents the linear interpolation of all data.}
    \label{fig14}
\end{figure}

Figure~\ref{fig14} illustrates the change in the ratio of the
observed [O III] line fluxes
$F(\lambda4959+\lambda5007)/F(\lambda4363)$ with time. The graph
shows a systematic increase in its value: in 75 years
$F(\lambda4959+\lambda5007)/F(\lambda4363)$ rose by a factor of
$\sim$4.


\begin{table}[h!]
    \centering
     \caption{$R=I(\lambda4959+\lambda5007)/I(\lambda4363)$ }
 \begin{tabular}{ccc}
\hline
Year&$R$&Source\\
\hline
1945&13.3&Aller (1951)\\
1966-1967&20.8&Kaler et al. (1976)\\
1971&24.9&Barker (1978)\\
1990&32.0&Hyung and Aller (1996)\\
1996&38.7&Kwitter et al. (2003)\\
2011-2019&51.1& This paper\\
2019-2020&52.1& This paper\\

  \hline
    \end{tabular}

    \label{tabl6}
\end{table}

We corrected the observed O$^{+2}$ line fluxes for extinction with
$c(\text{H}\beta)=1.15$;
$R=I(\lambda4959+\lambda5007)/I(\lambda4363)$ from the published
data and our own observations are given in Table~\ref{tabl6}. The
undoubted increase in $R$ suggests a change in the parameters of
the [O III] line formation region.

Suppose that the electron density did not change in the entire
time of spectroscopic observations. We will take
$N_e=5\times10^{5}$ cm$^{-3}$ for it from Hyung and Aller (1996).
Let us estimate $T_e$ for different epochs using the 5Level code
(De Robertis et al. 1987). We found the electron density to have
decreased from 22 000 K in 1945 to 10 650 K in 2020.
$T_e\approx13~000$ K was obtained for the epoch of observations by
Hyung and Aller (1996) in 1990. Whereas this results acceptable
values of the electron temperature for the last decades
(1990-2020), $T_e$ appears extremely high for the epoch of first
observations, and the electron density needs to be increased to
decrease $T_e$.

As has been shown above, the temperature of the central star has
remained constant over the last 40 years. Therefore, it is not yet
clear what is responsible for the significant changes in the
parameters of the inner PN region.

Since the parameters of the gaseous envelope of Hb 12 have been
determined repeatedly using diagnostic diagrams (see, in
particular, Hyung and Aller 1996; Kwitter et al. 2003) and since
the diagnostic ratios $F(\lambda
6548+\lambda6583)/{F(\lambda5755)}$ [N II],
$F(\lambda6717)/{F(\lambda6731)}$ [S II],
$F(\lambda7135+\lambda7751)/{F(\lambda5191)}$ [Ar~III] are
determined not quite reliably, here we will not apply this method.
However, in our paper we have measured for the first time the
Balmer jump for the nebula, which allowed $T_e$ to be estimated in
the hydrogen emission region.

The gas electron temperature can be calculated in accordance with
the formula (Liu et al. 2001)

\begin{center}
$T_{e}(BJ)=
368\times\left(1+0.259\frac{\mathrm{He}^{+}}{\mathrm{H}^{+}}+3.409
\frac{\mathrm{He}^{2+}}{\mathrm{H}^{+}}\right)
\times\left(\frac{BJ}{I(\mathrm{H} 11)}\right)^{-3/2}$,
\end{center}

where He$^{+}$/H$^{+}$ and He$^{2+}$/H$^{+}$ express the relative
abundances of neutral and ionized helium in the nebula,
respectively, $BJ$ is the Balmer jump defined as the difference of
the nebular continuum intensities before and after the jump, and
$I$(H11) denotes the intensity in the $\lambda$3771 hydrogen line.
For Hb 12 in the 2020 CMO spectrum we measured the Balmer jump to
be $BJ=7.06\times10^{-13}$ erg cm$^{-2}$ s$^{-1}$ \AA$^{-1}$. The
extinction-corrected line intensity I(H11) is 7.60$\times10^{-12}$
erg cm$^{-2}$ s$^{-1}$; the relative abundance of neutral helium
from the data of Hyung and Aller (1996) is
He$^{+}$/H$^{+}$=$7.04\times10^{-5}$, while the abundance of
ionized helium is negligible. Our calculation based on this
formula gives $T_{e}(BJ)\simeq12 980$ K.

\section*{CONCLUSIONS}

We presented the results of our spectroscopic observations of the
young compact PN Hb 12 in 2011--2020. We measured the absolute
H$\beta$ fluxes and derived the relative intensities of $\sim$50
nebular emission lines in the range $\lambda$3687-9532.

The extinction coefficient $c(\text{H}\beta)$ was found from
Balmer and Paschen hydrogen lines. It allowed us to correct the
data for interstellar reddening and to estimate the distance to
the nebula by analyzing the interstellar extinction maps for the
Galaxy.

Based on new and previously published data, we studied the
behavior of the H$\beta$ flux and the relative intensities of
individual lines in the time of spectroscopic observations of
Hb~12 from 1945 to 2020. We detected a number of probable trends,
with an enhancement of the relative intensities of nebular lines
and a weakening of the auroral O$^{+2}$ line being the most
significant one.

The rise in the relative intensity of the [O III] $\lambda$5007
line with time could be associated with an increase in the
temperature of the ionizing source. However, the behavior of the
relative intensity of the [N III] $\lambda$3869 line, which
exhibited no strengthening, but showed a falling trend, does not
support this hypothesis. Furthermore, the nebular core temperature
estimated by the energy balance method, $T_*\sim41~000$ K,
virtually coincides with the value obtained almost 40 years ago
(Preite-Martinez and Pottasch 1983).

Thus, we concluded that the variability of the spectrum for Hb 12
is attributable primarily to the variations in nebular parameters
rather than the evolution of the central star in the post-AGB
stage.

The ratio of the fluxes of the summary nebular O$^{+2}$ lines flux
to the auroral one, $R=I(\lambda5007+\lambda4959)/I(\lambda4363)$,
showed a linear trend, while its value increased by a factor of 4
from 1945 to 2020. The electron temperature was estimated for
different epochs. The value of $T_e$ was shown to decrease with
time. A decrease in the electron density in the [O~III] line
formation region is not ruled out either. What causes the change
in the parameters of the inner PN region remains an open question.

The electron temperature of the nebula in the H I line emission
region was determined from our measurements of the Balmer jump.

In addition, the CMO spectra were analyzed for the presence of
absorption features that could belong to the spectrum of the cool
companion of the central star. No traces of the companion were
detected.

\section*{ACKNOWLEDGMENTS}

The work was performed using the equipment purchased through the
funds of the Development Program of the Moscow State University.
In our studies we widely used the ADS, SIMBAD, and VIZIER
databases.

\bigskip
\section*{REFERENCES}
 \bigskip

\begin{enumerate}

\item A. Acker, B. Raytchev, J. Koeppen, and B. Stenholm, Astron.
Astrophys. Suppl. Ser. {\bf 89}, 237 (1991).

\item F. Ahern, Astrophys. J. {\bf 197}, 635 (1975).

\item F. Ahern, Astrophys. J. {\bf 223}, 901 (1978).

\item L. Aller, Astrophys. J. {\bf 113}, 125 (1951).

\item V. P. Arkhipova, M. A. Burlak, N. P. Ikonnikova, G. V.
Komissarova, V. F. Esipov, and V. I. Shenavrin, Astron. Lett. {\bf
46}, 100 (2020).

\item T. Barker, Astrophys. J. {\bf 219}, 914 (1978).

\item T. Bl\"{o}cker, Astron. Astrophys. {\bf 299}, 755 (1995).

\item A.G.A. Brown, A. Vallenari, T. Prusti, et al. (Gaia
Collab.), Astron. Astrophys. {\bf 616}, 10 (2018).

\item A.G.A. Brown, A. Vallenari, T. Prusti, J.H.J. de Bruijne, C.
Babusiaux, M. Biermann, O.L. Creevey, D.W. Evans, et al. (Gaia
Collab.), arXiv: 2012.01533 (2020).

\item M. A. Burlak and V. F. Esipov, Astron. Lett. {\bf 36}, 752
(2010).

\item J.H. Cahn, J.B. Kaler, and L. Stanghellini, Astron.
Atrophys. Supp. Ser. {\bf 94}, 399 (1992).

\item E.R. Capriotti, Astrophys. J. {\bf 140}, 632 (1964).

\item J.A. Cardelli, G.C. Clayton, and J.S. Mathis, Astrophys. J.
{\bf 345}, 245 (1989).

\item N. Chornay and N.A. Walton, arXiv:2102.13654v1 (2021).

\item D.M. Clark, J.A. L\'{o}pez, M.L. Edwards, and C. Winge,
Astron. J. {\bf 148}, 98 (2014).

\item J.E. Drew, R. Greimel, M.J. Irwin et al., MNRAS {\bf 362},
753 (2005).

\item A.V. Escudero, R.D.D. Costa, and W.J. Maciel, Astron.
Astrophys. {\bf 414}, 211 (2004).

\item M. Fa\'{u}ndez-Abans and W.J. Maciel, Astron. Astrophys.
{\bf 158}, 228 (1986).

\item D.J. Frew, Q.A. Parker, I.S. Boji\v{c}i\'{c}, MNRAS {\bf
455}, 1459 (2016).

\item C. Giammanco, S.E. Sale, R.L.M. Corradi et al., Astron.
Astrophys. {\bf 525}, A58 (2011).

\item T. Giannini, S. Antoniucci, B. Nisini, F. Bacciotti, and L.
Podio), Astrophys. J. {\bf 814}, 52 (2015).

\item I.N. Glushneva, V.T. Doroshenko, T.S. Fetisova, T.S.
Khruzina, E.A. Kolotilov, L.V. Mossakovskaya, S.L. Ovchinnikov,
and I.B.Voloshina, VizieR Online Data Catalog III/208 (1998).

\item I. Gonz\'{a}lez-Santamar\'\i a, M. Manteiga, A. Manchado, A.
Ulla, and C. Dafonte, Astron. Astrophys. {\bf 630}, A150 (2019).

\item G.M. Green, E. Schlafly, C. Zucker, J.S. Speagle, and D.
Finkbeiner, Astrophys. J. {\bf 887}, 93 (2019).

\item M. Hajduk, P.A.M. van Hoof, K. Gesicki, A. A. Zijlstra, S.
K. G\'{o}rny, and M. G\l adkowski, Astron. Astrophys. {\bf 567},
A15 (2014).

\item M. Hajduk, P.A.M. van Hoof, and A.J. Zijlstra, Astron.
Atrophys. {\bf 573}, A65 (2015).

\item C.H. Hsia, W.H. Ip, and J.Z. Li,  Astrophys. J. {\bf 131},
3040 (2006).

\item E. Hubble, PASP {\bf 33}, 174 (1921).

\item D.G. Hummer, and P.J. Storey, MNRAS {\bf 224}, 801 (1987).

\item S. Hyung and L. Aller, MNRAS {\bf 278}, 551 (1996).

\item B.W. Jiang, Ke Zhang, A. Li, and C.M. Lisse, Astrophys. J.
{\bf 765}, 72 (2013).

\item H.M. Johnson, B. Balick, and A.R. Thompson, Astrophys. J.
{\bf 233}, 919 (1979).

\item J.B. Kaler,  Astrophys. J. {\bf 220}, 887 (1978).

\item J.B. Kaler, L.H. Aller, and S.J. Czyzak, Astrophys. J. {\bf
203}, 636 (1976).

\item R.L. Kingsburg and M.J Barlow, MNRAS {\bf 257}, 317 (1992).

\item L.N. Kondratyeva, Astron. and Astrophys. Transactions {\bf
24}, 291 (2005).

\item E.B. Kostyakova and V.P. Arkhipova, Astron. Rep. {\bf 53},
1155 (2009).

\item K.B. Kwitter, R.B.C. Henry, and J.B. Milingo, Publ. Astron.
Soc. Pacific {\bf 115}, 80 (2003).

\item S. Kwok and C.H. Hsia, Astrophys. J. {\bf 660}, 341 (2007).

\item X.-W. Liu, M.J. Barlow, M. Cohen, I.J. Danziger, S.-G. Luo,
J.P. Baluteau, P. Cox, R.J. Emery, T. Lim, and D. Pequignot, MNRAS
{\bf 323}, 343 (2001).

\item K.L. Luhman and G.H. Rieke, Astrophys. J. {\bf 461}, 298
(1996).

\item O. De Marco, T.C. Hillwig, and A.J. Smith, Astron. J. {\bf
136}, 323 (2008).

\item M.M. Miller Bertolami, Astron. Atrophys. {\bf 588}, A25
(2016).

\item L.F. Miranda, and J. Solf, Astron. Atrophys. {\bf 244}, 353
(1989).

\item C.R. O'Dell, Astrophys. J. {\bf 138}, 293 (1963).

\item R. Ohsawa, T. Onaka, I. Sakon, M. Matsuura, and H. Kaneda,
Astron. J. {\bf 151}, 93 (2016).

\item M. Peimbert, in Planetary Nebulae, Observation and Theory,
ed. Y., Terzian (Dordrecht: Reidel), IAU Symp., {\bf 76}, 215
(1978).

\item M. Perinotto, Astrophys. J. Suppl. Ser. {\bf 76}, 687
(1991).

\item J.P. Phillips, Astrophys. J. Suppl. Ser. {\bf 139}, 199
(2002).

\item J.P. Phillips, MNRAS {\bf 353}, 589 (2004).

\item A.J. Pickles, Astrophys. J. Suppl. Ser. {\bf 59}, 33 (1985).

\item S. A. Potanin, A. A. Belinskii, A. V. Dodin, et al., Astron.
Lett. {\bf 46}, 837 (2020).

\item S. R. Pottasch, Planetary Nebulae (Kluwer, Dordrecht, 1984).

\item A. Preite-Martinez and S.R. Pottasch, Astron. Astrophys.
{\bf 126}, 31 (1983).

\item C. Quireza, H.J. Rocha-Pinto, and W.J. Maciel, Astron.
Astrophys. {\bf 475}, 217 (2007).

\item M.M. De Robertis, R.J. Dufour, and R.W. Hunt, JRASC {\bf 81}
(1987).

\item R.J. Rudy, G.S. Rossano, P. Erwin, R.C. Puetter, and W.A.
Feibelman, Astrophys. J. {\bf 105}, 1002 (1993).

\item M.J. Seaton, MNRAS {\bf 187}, 73 (1979).

\item S.G. Sergeev and F. Heisberger, A Users Manual for SPE. Wien
(1993).

\item R. Tylenda, A. Acker, and B. Stenholm, Astron. Astrophys.
Suppl. Ser. {\bf 102}, 595 (1993).

\item E. Vassiliadis and P.R. Wood, Astrophys. J. Suppl. Ser. {\bf
92}, 125 (1994).

\item N.M.H. Vaytet, A.P. Rushton, M. Lloyd, J.A. L\'{o}pez, J.
Meaburn, T.J. O'Brien, D.L. Mitchell, and D. Pollacco, MNRAS {\bf
398}, 385 (2009).

\item C.Y. Zhang, Astrophys. J. {\bf 410}, 239 (1993).

\item C.Y. Zhang and S. Kwok, Astron. Astrophys. {\bf 237}, 479
(1991).

\end{enumerate}

Translated by V. Astakhov
\newpage

\section*{Appendix: Spectroscopic Data for Hb 12}

\begin{table}[h!]
    \scriptsize
    \centering
    \caption{The observed emission line intensities expressed in units of $F(\text{H}\beta)=100$ that were derived during the CAS
observations in 2011--2019}
    \begin{tabular}{|c|c|c|c|c|c|c|c|c|c|}
    \hline
     $\lambda$, \AA & Species & F(11) & F(15) & F(16) & F(17) & F(18) & F(19) & F(19a) & F(19b) \\
    \hline
    4101 & H I      & 12.3  & 18.3  & 14.2  & -     & 15.8  & 11.3  &   -  &   -  \\
    4340 & H I      & 33.9  & 33.2  & 34.1  & 37.2  & 34.0  & 31.1  & 39.1 &34.2  \\
    4363 & [O III]  & 10.0  & 9.6   & 9.4   & 10.4  & 9.5   & 9.18  &12.4  &8.4   \\
    4471 & He I     & 3.68  & 4.08  & 3.41  & 3.40  & 3.37  & 3.91  &      &      \\
    4387 & He I     & -     & -     &    -  & -     & -     &  -    & 0.69 & -    \\
    4713 & He I     & 0.66  & 0.71  & 0.76  & -     & -     & 1.59  & 0.96 &0.91  \\
    4861 & H I      & 100   & 100   & 100   & 100   & 100   & 100   &100   &100   \\
    4920 & He I  & -     & -     &  -    &  -    &  -    & -     &2.0   & 0.99 \\
    4959 & [O III]  & 183   & 181   & 189   & 181   & 186   & 203   &183   &167   \\
    5007 & [O III]  & 574   & 564   & 586   & 571   & 584   & 629   &570   &522   \\
    5147 & [Fe IV]  & -     & -     & -     & -     & -     & 0.14  &-     &-     \\
    5159 & [Fe II]  & -     & -     & -     & -     & -     & 0.12  &-     &-     \\
    5192 & [Ar III] & 0.44  & 0.27  & -     & -     & 0.37  & 0.60  &0.47  &0.31  \\
    5270 & [Fe III] & 0.37  & -     & -     & -     & -     & 0.61  &0.38  &0.44  \\
    5517 & [Cl III] & -     & -     & -     & -     & 0.16  & -     & -    & -    \\
    5537 & [Cl III] & -     & 0.26  & 0.29  & -     & 0.40  & 0.30  &0.36  &0.24  \\
    5577 & [O I]    & 0.89  & 0.39  & 0.43  & 0.78  & -     & 0.42  &0.16  &0.30  \\
    5666 & N II     & -     & -     & 0.16  & -     & -     & -     & -    &-     \\
    5680 & N II     & -     & -     & 0.28  & -     & -     & 0.17  &0.59  &0.29  \\
    5755 & [N II]   & 10.7  & 9.15  & 11.7  & 9.67  & 9.70  & 8.65  & 11.2 &8.42  \\
    5876 & He I     & 28.0  & 27.4  & 31.5  & 26.8  & 30.3  & 24.2  & 29.3 &23.2  \\
    5959 & O I      & -     & 0.23  & 0.35  & -     & 0.21  & 0.19  & 0.43 &0.24  \\
    5979 & Si II    & -     & 0.18  & 0.27  & -     & 0.20  & 0.24  & 0.39 &0.17  \\
    6048 & O I      & -     & 0.31  & 0.40  & 0.26  & 0.26  & 0.23  & 0.38 &0.24  \\
    6300 & [O I]    & 4.63  & 2.43  & 2.91  & 2.93  & 3.29  & 2.47  &2.91  &2.45  \\
    6312 & [S III]  & 8.02  & 8.56  & 8.40  & 7.88  & 8.55  & 7.69  &9.54  &7.48  \\
    6347 & Si II    & -     & 0.39  & 0.45  & 0.32  & -     & 0.33  &0.43  &0.27  \\
    6363 & [O I]    & 0.77  & 0.99  & 1.20  & 0.77  & 0.92  & 0.70  &1.40  &0.81  \\
    6563 & H I      & 662   & 690   & 671   & 671   & 683   & -     &641   &626   \\
    6584 & [N II]   & 65.0  & 71.0  & 95.0  & 79.0  & 80.0  & 84.5  &73.2  &62.8  \\
    6678 & He I     & 9.98  & 9.05  & 10.53 & 8.80  & 9.60  & 8.67  &10.8  &7.98  \\
    6717 & [S II]   & 0.62  & 0.55  & 0.64  & 0.57  & 0.99  & 0.43  &0.58  &0.77  \\
    6731 & [S II]   & 1.21  & 1.17  & 1.29  & 1.28  & 1.24  & 0.96  &1.62  &1.46  \\
    7002 & O I      & -     & -     & 0.25  & -     & -     & 0.23  &0.58  &0.22  \\
    7065 & He I     & 34.73 & 33.12 & 33.35 & 36.65 & 32.70 & 23.5  &39.2  &26.7  \\
    7135 & [Ar III] & 63.07 & 60.27 & 60.18 & 64.11 & 60.29 & 43.1  &69.0  &47.8  \\
    7236 & C II     & -     & -     & 0.85  & -     & 0.85  & 0.46  &-     &0.74  \\
    7253 & O I      & 0.76  & 0.52  & 0.52  & -     & 0.85  & 0.40  &0.61  &0.70  \\
    7281 & He I     & 3.45  & 2.97  &  2.97 & -     & 5.08  & 3.14  &3.48  &2.55  \\
    7325 & [O II]   & 126   & 97.56 & 98.1  & 98    & 97    & 71.0  &120.1 &79.9  \\
    7376 & ?     & -     & -     &  -    & -     &-      &-      & 0.26 & -    \\
    7467 & N I?        & -     & -     &  -    & -     &-      &-      &0.23  & -    \\
    7751 & [Ar III] & 21.06 & 15.63 & 15.63 & 19.51 & 17.6  & 11.3  &17.4  &-     \\
    8323 & H I      & -     & 0.06  & 0.11  & 0.12  & 0.20  & 0.15  &-     &-     \\
    8334 & H I      & -     & 0.23  & 0.20  & 0.24  & 0.28  & 0.24  &-     &-     \\
    8346 & H I      & -     & 0.50  & 0.58  & 0.54  & 0.44  & 0.55  &-     &-     \\
    8361 & He I     & 1.17  & 0.93  & 1.03  & 0.96  & 1.10  & 0.96  &-     &-     \\
    8375 & H I      & 0.68  & 0.61  & 0.68  & 0.57  & 0.57  & 0.67  &-     &-     \\
    8392 & H I      & 1.04  & 0.94  & 1.05  & 0.88  & 0.74  & 0.81  &-     &-     \\
    8413 & H I      & 1.08  & 1.09  & 1.09  & 1.11  & 1.09  & 0.84  &-     & -    \\
    8442 & H I/O I  & 10.4  & 8.45  & 7.63  & 7.77  & 9.63  & 7.05  &10.2  & -    \\
    8467 & H I      & 2.19  & 1.77  & 1.42  & 1.56  & 1.71  & 1.44  &-  & -    \\
    8502 & H I      & 2.87  & 2.24  & 2.19  & 1.95  & 2.22  & 2.02  &2.82  & -    \\
    8545 & H I      & 3.22  & 2.41  & 2.30  & 2.17  & 2.67  & 2.53  &3.12  &-     \\
    8598 & H I      & 3.65  & 3.11  & 2.35  & 2.59  & 3.34  & 2.43  &4.16  & -    \\
    8665 & H I      & 5.40  & 4.19  & 5.11  & 3.82  & 4.33  & 3.36  &3.50  &  -   \\
    8750 & H I      & 6.32  & 4.51  & 4.06  & 4.57  & 5.20  & 3.64  &5.27  &   -  \\
    8862 & H I      & 8.19  & 5.65  & 5.36  & 6.07  & -     & 4.51  &6.02  &    - \\
    9015 & H I      & 10.9  & -     & 7.05  & 7.31  & -     & 5.24  & 7.89 &-     \\
    9069 & [S III]  & 119   & -     & -     & 82.7  & 102   & -     & 92.9 & -    \\
    9229 & H I      & -     & -     & -     & 11.64 & -     & -     & -    &  -   \\
    9532 & [S III]  & -     & -     & -     & 198   & -     & -     &  -   &   -  \\
    \hline
    \end{tabular}
\end{table}

\begin{table}[h!]
    \small
    \centering
    \caption{The observed emission line intensities expressed in units of $F(\text{H}\beta)=100$ that were derived during the
observations at the 2.5-m CMO telescope on November 7, 2019, and
January 20, 2020}
    \begin{tabular}{|c|c|c|c||c|c|c|c|}
\hline
$\lambda$, \AA & Species & F(19c) & F(20) & $\lambda$, \AA & Species & F(19c) & F(20) \\
\hline

3687 & H I          & -    & 0.54 &5273 & [Fe III]     & 0.31 & 0.23\\
3692 & H I          & -    & 0.60 &5299 & O I & -&0.10\\
3697 & H I          & -    & 0.67 &5513 & O I & -    &0.07\\
3705 & He I         & 0.75 & 1.20 &5517 & [Cl III] & -&0.08 \\
3712 & H I          & 0.74 & 1.05 &5537 & [Cl III]     & -    & 0.30 \\
3726 & [SIII]/[OII] & 6.12 & 5.75 &5555 & O I          & -    & 0.14 \\
3734 & H I          & 1.16 & 1.41 &5577 & [O I]        & -    & 0.12\\
3750 & H I          & -    & 1.89 &5667 & N II         & -    & 0.32 \\
3771 & H I          & -    & 2.34 &5680 & N II         & -    & 0.19  \\
3798 & H I          & -    & 3.06 &5755 & [N II]       & 10.5 & 12.1\\
3819 & He I         & 1.07 & 0.84 &5868 & Si II        & -    & 0.04  \\
3835 & HI/HeI       & 4.02 & 4.04 &5876 & He I         & 26.8 & 31.6  \\
3869 & [Ne III]     & 31.8 & 33.9 &5932 & N II         & -    & 0.12\\
3889 & H I          & 7.01 & 7.33 &5942 & N II         & -    & 0.20\\
3967 & [Ne III]     & 18.2 & 20.6 &5958 & O I          & 0.30 & -     \\
4009 & [NeIII]/HI   & -    & 0.19 &5978 & Si II        & 0.28 & 0.23 \\
4026 & He I         & -    & 1.68 &6048 & O I          & 0.58 & 0.33\\
4069 & [S II]       & 1.16 & 0.97 &6300 & [O I]        & 0.86 & 2.69  \\
4076 & [S II]       & -    & 0.33 &6312 & [S III]      & 8.97 & 10.6\\
4101 & H I          & 15.3 & 15.4 &6347 & Si II        & 0.42 & 0.42  \\
4121 & He I         & 0.32 &  0.28&6363 & [O I]        & 0.86 & 0.91\\
4144 & He I         & 0.29 &0.24  &6371 & Si II        & 0.30 & 0.26  \\
4340 & H I          & 35.4 &33.3  &6461 & N II         & 0.11 & 0.07\\
4363 & [O III]      & 10.5 &10.1  &6481 & N II         & 0.11 & 0.10 \\
4388 & He I         & 0.57 &0.53  &6548 & [N II]       & 21.2 & 19.9 \\
4414 & [Fe II]      & -    &0.26  &6563 & H I          &  -   & 881\\
4471 & He I         & 4.49 &4.17  &6584 & [N II]       & 65.9 & 61.3   \\
4631 & N II         & -    & 0.15 &6678 & He I         & 11.0 & 11.8\\
4641 & N III        & -    & 0.34 &6717 & [S II]       & 0.65 & 0.44 \\
4649 & O II         & -    & 0.31 &6731 & [S II]       & 1.31 & 0.88 \\
4676 & O II         & -    & 0.04 &7002 & O I          & 0.48 & 0.39\\
4686 & He II        & -    & 0.14 &7065 & He I         & 38.7 & 39.0 \\
4701 & [Fe III]     & -    & 0.24 &7136 & [Ar III]     & 62.6 & 67.0\\
4713 & He I         & 1.13 & 0.93 &7155 & [Fe II]      & -    & 0.10  \\
4740 & [Ar IV]      & 0.19 & 0.22 &7161 & He I         & -    & 0.11 \\
4861 & H I          & 100  & 100  &7172 & [Ar IV]      & 0.17 & 0.16  \\
4921 & He I         & 1.58 & 1.49 &7232 & C II         & -    & 0.36 \\
4932 & [O III]      & -    & 0.15 &7236 & [Ar IV]      & 1.44 & 0.75\\
4959 & [O III]      & 186  & 188  &7254 & O I          & 1.10 & 0.98\\
5007 & [O III]      & 598  & 601  &7281 & He I         & 2.76 & 3.26 \\
5047 & Si II        & 0.28 & 0.29 &7307 & O III?       & 1.05 & 1.13\\
5056 & Si II        & 0.25 & 0.22 &7319 & [O II]       & 57.4 & 65.1 \\
5147 & [Fe IV]      & 0.22 & 0.18 &7330 & [O II]       & 46.3 & 52.6\\
5191 & [ArIII]/[NI] & 0.21 & 0.25 &7377 & [Ni II]      & -    & 0.08\\
5197 & [N I]        & -    & 0.12 &7444 & N II         & -    & 0.13 \\
5263 & [Fe II]      & 0.16 & 0.08&&&& \\

\hline
\end{tabular}
\end{table}

\end{document}